# The future of hearing aid technology
Can technology turn us into superheroes?

Volker Hohmann[1,2,3]

[1] Department of Medical Physics and Acoustics, University of Oldenburg, Oldenburg, Germany

[2] Hörzentrum Oldenburg gGmbH, Oldenburg, Germany

[3] Cluster of Excellence Hearing4all Oldenburg, Germany

**Abstract:**

**Background.** Hearing aid technology has proven successful in the rehabilitation of hearing loss, but its performance is still limited in difficult everyday conditions characterized by noise and reverberation.

**Objectives.** Introduction to the current state of hearing aid technology and presentation of the current state of research and future development.

**Methods.** Current literature is analyzed and several specific new developments are presented.

**Results.** Both objective and subjective data from empirical studies show the limitation of current technology. Examples of current research show the potential of machine-learning based algorithms and multi-modal signal processing for improving speech processing and perception, of using virtual reality for improving hearing device fitting and of mobile health technology for improving hearing-health services.

**Conclusions.** Hearing device technology will remain a key factor in the rehabilitation of hearing impairment. New technology such as machine learning, and multi-modal signal processing, virtual reality and mobile health technology will improve speech enhancement, individual fitting and communication training.

**Keywords:**

Hearing Acoustics; Hearing Impairment; Hearing Aids; Hearing Devices; Speech Processing; Speech Enhancement; Machine Learning; Virtual Reality; Mobile Health Technology

Ubiquitous smartphone technology offers impressive services and has changed the life of millions of users. Smartphones are increasingly being used as a basis for mobile health applications, which are of broad interest for research and development and will improve personalized healthcare in the future ([1]). First services to support smartphone users with hearing impairment have been deployed, such as, e.g., automatic speech transcription, personalized sound and noise reduction as well as mobile-health supported hearing health care services ([2],[3]). These impressive developments help hearing-impaired people cope with their handicap, but acoustic communication at home, in public and at the workplace, requires hearing devices to work "at the ear" of the user, with high sound quality, good speech perception, high listening comfort and good usability ([4]).

## Current state of hearing device technology

Hearing devices ([5],[6],[7],[8],[9]) pick up the sound in the environment of the user, process it and deliver processed sound to the user by different means of stimulation, depending on the type of hearing loss: Acoustic (hearing aids and hearables), electric (cochlear implants) or mechanic (bone-conduction devices or devices with direct stimulation of the temporal bone). For the most common case of mild or moderate sensorineural hearing loss, hearing aids are mainly placed behind the ear (BTE). The lower right part of Fig. 1 shows a typical BTE device with two microphones in its shell and a loudspeaker (called "receiver") placed at the entrance of the ear canal for acoustic stimulation.

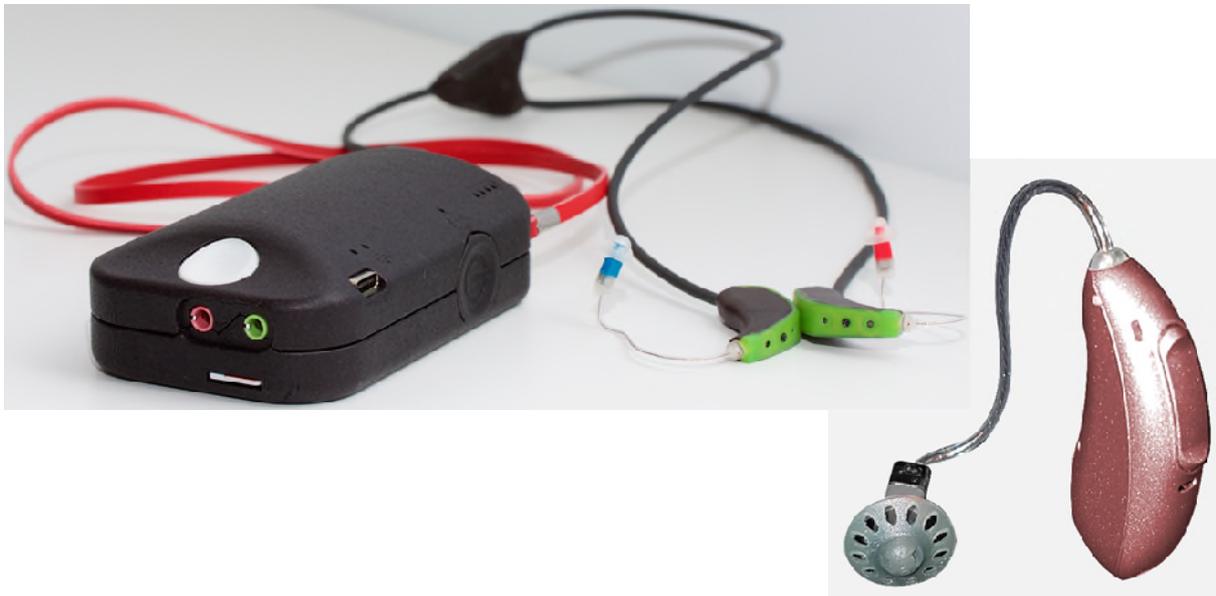

*Figure 1: Lower right: Commercial Behind-the-Ear (BTE) hearing aid, with the receiver in the ear canal (RIC). The shell is placed behind the ear, and a soft dome holds the receiver in place in the ear canal (Kevin Pollak, CC BY 3.0). Upper left: The Portable Hearing Lab (PHL) research hearing aid, employing two high-quality BTE devices and a freely programmable binaural signal processor (Hendrik Kayser, CC BY-NC-ND 4.0).*

Being very small, BTE devices hold a tiny battery with a supply voltage of just over 1 Volt. Yet, they employ sound processors with high computational power, have wireless connectivity options, including Bluetooth low-energy (BLE) and a wireless link between devices at left and right ear for binaural processing, deliver high levels of sound and run on a single battery for a couple of days (or a day, for rechargeable batteries). Because of the high integration and efficiency, sound processors are not freely programmable, i.e., commercial hearing devices come with a set of fixed processing options. Only the processing parameters can be modified to activate or deactivate processing options, e.g., switching noise reduction on or off, and to fit the processing to the individual hearing loss in terms of amplification, frequency response and other factors. To enable basic research into novel algorithms and fitting methods, research devices are required, which are freely programmable and offer the same electro-acoustic properties as commercial hearing aids. One example is the Portable Hearing Laboratory (PHL, [10][11]) shown in the upper left part of Fig. 1. The PHL is being used at the author's and a several other labs worldwide for hearing-aid research towards novel devices and improved hearing-aid performance.

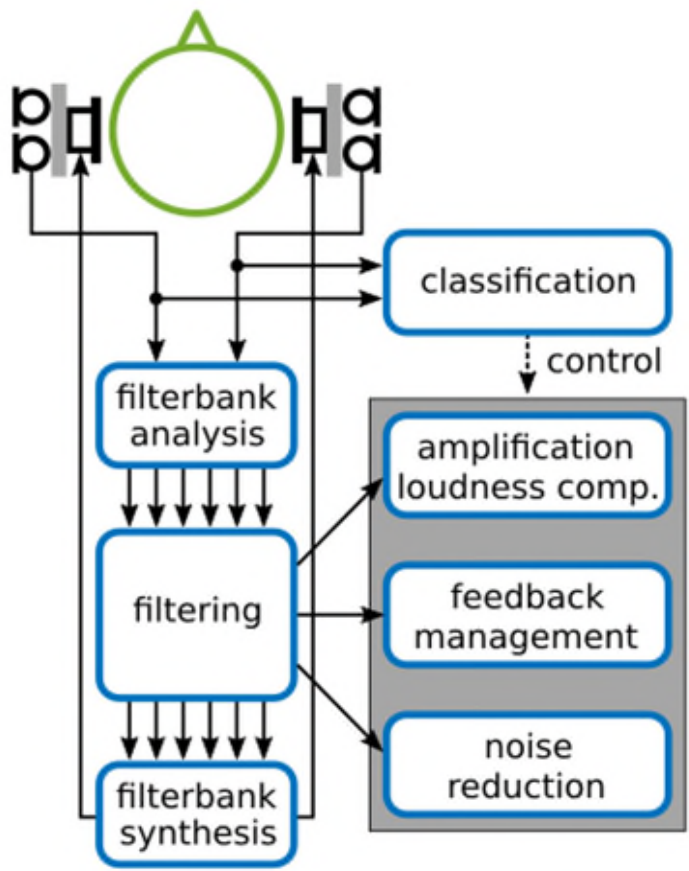

*Figure 2: Signal processing block diagram of a generic binaural hearing aid.*

Fig. 2 shows the block diagram of a standard hearing aid sound processing scheme, which is implemented in a similar form in commercial hearing devices and in research devices such as the PHL. Sound is picked up by one or two microphones at each ear and then digitized for digital sound processing. The signals are usually transformed to the frequency domain by, e.g., Fourier transform or by a bank of frequency filters, to enable frequency-specific processing. Processing usually comprises several blocks:

- Frequency-specific amplification and dynamic compression to compensate for an increased threshold of hearing while not overamplifying high-level (i.e., loud) sounds.
- Noise reduction to improve speech perception in noisy and reverberant environments.
- Feedback management to suppress distortion and howling due to the acoustic and mechanic feedback of amplified sound from the receiver to the microphones. In particular, feedback easily occurs in open fittings were the ear canal is not acoustically sealed by the device, requiring advanced feedback management.

Finally, processed sound is transformed back to the time domain by adding up all frequency specific signals, converted to analog and delivered to the receiver. Further processing options include classification schemes, which aim at adapting the processing dependent on the detected acoustic condition, e.g., automatically switching on noise reduction when speech-in-noise is detected, or changing the amplification settings when music is detected. More detailed information on sound processing in hearing devices can be found in literature ([5],[6],[7],[8],[9]).

A recent extensive survey ([4]) shows that hearing aids improved over the last decades and that performance levels in terms of user satisfaction and perceived benefit are rather high, in particular for devices from 2015 or later. Yet, studies show that performance of recent hearing aids in noisy conditions is high when tested in controlled laboratory conditions, but remains limited in real-life conditions ([12]) even for premium devices. A study using virtual reality to better simulate real-life acoustic communication showed that head movements and non-stationarity of noise signals are relevant factors that may explain this discrepancy between laboratory and real-life performance results ([13]). In any case, data show that there is an urgent need for for hearing devices to work better in noisy and reverberant acoustic conditions. In addition, a meta-study looking into reasons for non-use of hearing aids ([14]) identified insufficient comfort related to wearing the hearing aid as a factor, which includes unpleasant sound, sound being too loud, and limited benefit overall. To overcome these limitations, authors suggest research into the "fit and comfort of the hearing aid". To this end, ongoing research focuses on better fitting methods (e.g., [15][16]) and hearing or communication training (e.g., [17]).

## Requirements for future technology

*The communication loop*

Current hearing devices have been designed assuming a passive model of communication, i.e., assuming that a sender emits a sound, which is transmitted to and perceived by a listener without any interaction between listener and sender and with a static sound transmission path between sender and receiver. Therefore, most established laboratory methods to test hearing aids use a couple of loudspeakers at fixed locations for speech and noise presentation and hearing aids attached to a dummy head positioned at a fixed location relative to the loudspeakers. In such conditions, current hearing aids prove to be very effective in suppressing noise and enhancing speech. In reality, however, most communication evolves in a dynamic loop between sender (or senders) and listener, involving active behavior such as turn-taking, relative movement between senders and listener, as well as other active behavior such as turning of the head. This means that the passive communication model is not appropriate and must be replaced by a *communication loop model* to better reflect the dynamics of interactive real-life communication. In particular, the hearing aid needs to become aware of the evolving communication loop to be able to adapt its signal processing to the dynamically changing communication condition. A particular problem in integrating the communication loop into the device is selective attention.

*The problem of selective attention*

When communicating in a difficult communication condition with several sound sources and background noise, human listeners actively focus on a certain sound source they want to listen to, while putting all other sound sources in the background. This focus of attention can be voluntarily shifted to another source, e.g., in a conversation at a table with several people speaking at the same time. To support a listener in focusing their attention on a specific sound source, a hearing aid needs to know which of the currently active sound sources is in the focus of attention of the listener in order to be able to selectively enhance it. The problem is, that the focus of attention cannot be inferred easily from the acoustic signal alone. The next section will outline potential solutions to the selective attention problem.

## New developments

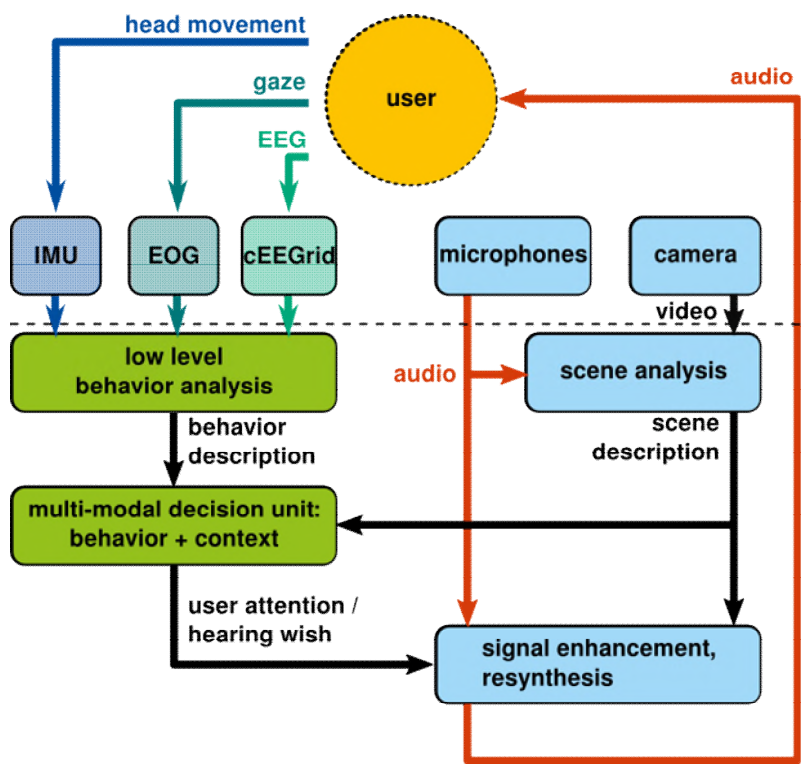

*Figure 3: Block diagram of the Immersive Hearing Device concept, which uses data from biosensors to estimate the user's focus of attention.*

Figure 3 shows a block diagram of the Immersive Hearing Device, a new hearing aid concept, which is under investigation at several labs worldwide and may solve the selective attention problem in the future. The general idea is to measure user behavior and activity using a multi-modal signal processing approach. To this end, different sensors attached to the hearing aid (left column of processing blocks in Figure 3), in particular an inertial measurement unit (IMU) to measure head movement, a set of electrodes close to the ear to measure eye gaze and potentially also several more electrodes around the ear (cEEGrid) to measure brain activity (EEG). The acoustic scene is analyzed from the microphone signals, estimating the presence, activity and spatial location of sound sources (right column of processing blocks). The scene analysis may be improved by visual data from a small camera attached to the hearing aid, e.g., by detecting mouth/lip movements ([18]). Next, a decision unit combines the acoustic scene data with the sensor data to estimate user attention. The decision unit uses machine learning techniques to learn the relation between acoustic scene and sensor data from large amounts of test data with the aim to reach a high level of accuracy in estimating the attended source. Finally, a signal enhancement block enhances the attended signal and presents it to the user. Currently, machine-learning based signal enhancement schemes are under development (e.g., [19][20]), which yield and improved signal enhancement and better sound quality than traditional methods. All processing blocks of the concept shown in Figure 4 are still under investigation, but first complete systems have already been demonstrated, e.g., gaze-based attention steering (Figure 4, [21]) and neuro-steered hearing devices (Figure 5, [22][23]).

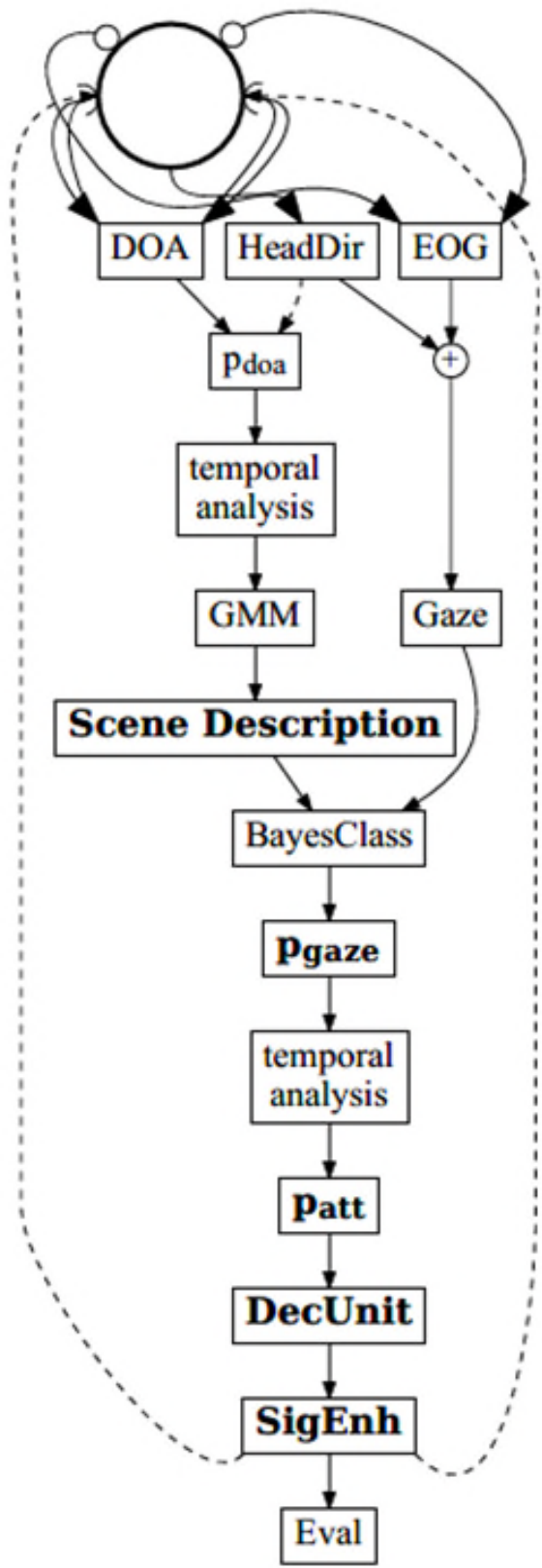

*Figure 4: Block diagram of a gaze-steered binaural speech enhancement scheme, implementing a first version of the Immersive Hearing Device concept (Figure adapted from [21]).*

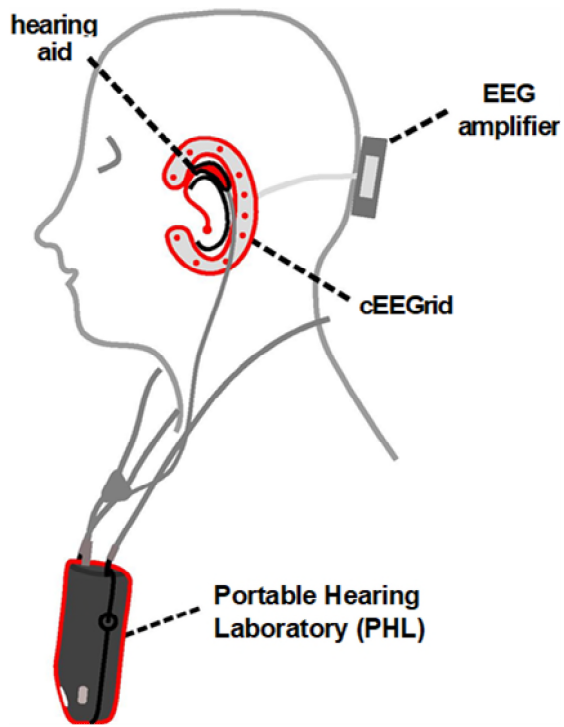

*Figure 5: Concept of a research version of a neuro-steered hearing device, which integrates mobile EEG (cEEGrid sensor) and the PHL research hearing aid to jointly process audio and EEG data (Figure replotted from [23]).*

*Virtual reality*

The Immersive Hearing Device closes the loop between the behaving listener and the acoustic environment. It cannot be tested in the established static laboratory environments, because natural behavior (in terms of, e.g., eye and head movement or brain activity) can only be achieved in realistic interactive communication environments. The implementation of these closed-loop conditions in the lab became possible with the advent of affordable virtual reality (VR) technology. Several labs worldwide set up interactive VR environments for the development and test of novel hearing devices ([24][25]). Figure 6 shows a photo of an example VR lab, the *Gesture Lab* at the University of Oldenburg ([24][26]).

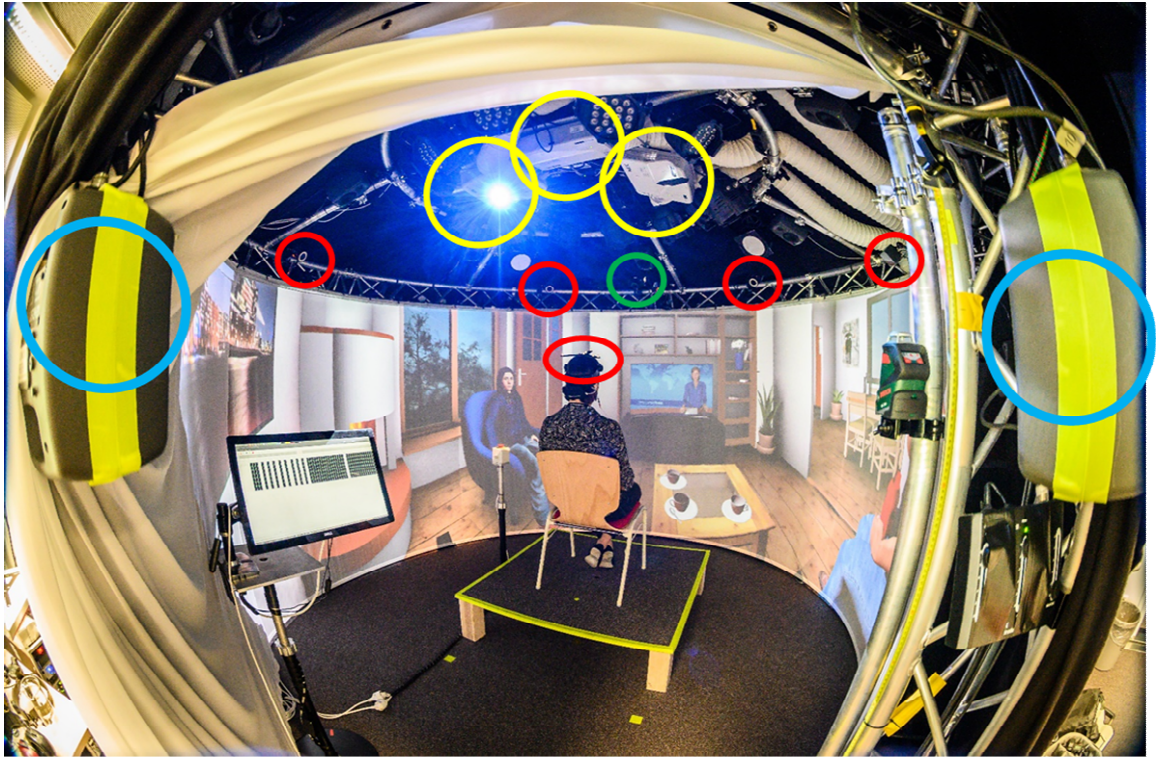

*Figure 6: Virtual Reality Lab at Oldenburg University, enabling audio-visual subject-in-the loop communication studies (photo: Hörtech gGmbH, CC BY-NC-ND 4.0).*

Due to the success of VR in hearing aid research and development, several researchers started investigating the use of VR for hearing-aid fitting and listening training. First results (e.g., [27][28][29][30]) show the feasibility of the approach, suggesting that VR will become an integral part of the hearing aid provision and fitting process in the future.

*Mobile health technology*

One important trend in in mobile technology is the use of earpieces in combination with mobile phones. Some of them, e.g., Apple AirPods, Samsung Galaxy Buds Pro or the IQbuds MAX, enable a hear-through mode, i.e., the sound in the environment is picked up by microphones in the earpiece, processed and played back in enhanced form through the earpiece, just like in hearing aids. These so-called Enhancing Personal Sound Amplification Product (PSAP) or hearables ([31]) are not medically regulated and are designed to support hearing in users with normal hearing. However, they also provide some amplification and frequency shaping in addition to speech enhancement and noise reduction. Thus, they may also support people with mild hearing losses, forming an entry point into hearing aid provision when the hearing loss grows over time and with age ([31][32]). Recently, a new medically regulated product category was introduced, so-called over-the-counter hearing device (OTC), which provide more amplification than a PSAP and may be used for mild to moderate hearing losses ([33][34]). For larger hearing losses, for people requiring more personal service, e.g., by audiologists or hearing-aid dispensers, and for patients requiring electric or mechanic stimulation, full medically regulated hearing devices are used. Such devices developed in recent years in terms of signal processing capabilities, as outlined above, but also adopted mobile technology and wireless mobile phone links to improve usability, e.g., device control, music streaming, self- and/or remote fitting and processing program selection ([35]). Furthermore, novel services were added for the target group of older hearing-device users such as, e.g., fall detection ([36]). The new product categories, PSAPs and OTCs as well as the integration with mobile phone technology are important factors in improving the rehabilitation of hearing loss for a broad range of patients and users.

## Summary and conclusions

Despite the tremendous improvement of and benefit from hearing aids since the introduction of all-digital signal processing in these devices in 1996, there are still problems with hearing in noise, the individual fitting of the device and its handling. Recent progress in research and development shows that new technology will improve the rehabilitation with hearing devices, but whether or not this will turn us into superheroes remains to be seen, as this term would mean that the hearing-impaired with a hearing device would perform better than a normal-hearing person without such a device.

In particular, the following conclusions can be drawn from the material presented above:

- Current hearing device technology is successful in the rehabilitation of hearing loss for less complex listening conditions, i.e., at rather high signal-to-noise ratios.
- For complex listening conditions, in which the target speech is superposed at a rather low signal-to-noise ratio with other sounds and is corrupted by reverberation, current speech enhancement schemes fall short of the performance of the normal-hearing auditory system. Hearing aid users therefore fail to guide their attention, to fully perceive speech and to achieve acoustic communication.
- Key for future technology is therefore (i) the estimation of the attended sound source, i.e., the sound source of a mixture of sounds that is currently in the focus of attention of the listener, and (ii) the selective enhancement of the attended source.
- First attempts towards that goal have been proven promising, but still lack full applicability.
- Literature shows that Machine Learning methods will be needed in addition to traditional multi-modal signal processing techniques to solve the problem.
- Hearing device fitting and hearing training using virtual reality is a promising technique to better individualize hearing aids and to improve individual hearing success with hearing devices.
- Combination of hearing aids with smartphone-based mobile health technology may contribute to an integrated approach to hearing health care.

## Acknowledgements

Funded by the Deutsche Forschungsgemeinschaft (DFG, German Research Foundation) – Project-ID 352015383 – SFB 1330 – project B1. Many thanks to G. Grimm, H. Kayser and all other members of the Auditory Signal Processing group.

## References

[1] Fan, K., & Zhao, Y. (2022). Mobile health technology: a novel tool in chronic disease management. Intelligent Medicine, 2(1), 41-47.

[2] Slaney, M., Lyon, R. F., Garcia, R., Kemler, B., Gnegy, C., Wilson, K., ... & Cerf, V. G. (2020). Auditory measures for the next billion users. Ear and hearing, 41, 131S-139S.

[3] Frisby, C., Eikelboom, R., Mahomed-Asmail, F., Kuper, H., & Swanepoel, D. W. (2022). m-Health Applications for Hearing Loss: A Scoping Review. Telemedicine and e-Health, 28(8), 1090-1099.

[4] Picou, E. M. (2020, February). MarkeTrak 10 (MT10) survey results demonstrate high satisfaction with and benefits from hearing aids. In Seminars in Hearing (Vol. 41, No. 01, pp. 021-036). Thieme Medical Publishers.

[5] Dillon H (2012). Hearing Aids. Thieme.

[6] Hohmann, V. (2008). Signal processing in hearing aids. In Handbook of Signal Processing in Acoustics (pp. 205-212). Springer, New York, NY.

[7] Kates J (2008) Digital Hearing Aids, Plural Publishing, San Diego, Calif, USA.

[8] Kollmeier, B., & Kiessling, J. (2018). Functionality of hearing aids: State-of-the-art and future model-based solutions. International journal of audiology, 57(sup3), S3-S28.

[9] Schaub, A. (2008). Digital hearing aids. Thieme.

[10] Kayser, H., Herzke, T., Maanen, P., Zimmermann, M., Grimm, G., & Hohmann, V. (2022). Open community platform for hearing aid algorithm research: open Master Hearing Aid (openMHA). SoftwareX, 17, 100953.

[11] Pavlovic, C., Kassayan, R., Prakash, S. R., Kayser, H., Hohmann, V., & Atamaniuk, A. (2019). A high-fidelity multi-channel portable platform for development of novel algorithms for assistive listening wearables. The Journal of the Acoustical Society of America, 146(4), 2878-2878.

[12] Wu, Y. H., Stangl, E., Chipara, O., Hasan, S. S., DeVries, S., & Oleson, J. (2019). Efficacy and effectiveness of advanced hearing aid directional and noise reduction technologies for older adults with mild to moderate hearing loss. Ear and Hearing, 40(4), 805.

[13] Hendrikse, M. M., Grimm, G., & Hohmann, V. (2020). Evaluation of the influence of head movement on hearing aid algorithm performance using acoustic simulations. Trends in hearing, 24, 2331216520916682.

[14] McCormack, A., & Fortnum, H. (2013). Why do people fitted with hearing aids not wear them?. International journal of audiology, 52(5), 360-368.

[15]     Fontan, L., Le Coz, M., Azzopardi, C., Stone, M. A., & Füllgrabe, C. (2020). Improving hearing-aid gains based on automatic speech recognition. The Journal of the Acoustical Society of America, 148(3), EL227-EL233.

[16]     Oetting, D., Bach, J. H., Krueger, M., Vormann, M., Schulte, M., & Meis, M. (2019). Subjective loudness ratings of vehicle noise with the hearing aid fitting methods NAL-NL2 and trueLOUDNESS. In Proceedings of the International Symposium on Auditory and Audiological Research (Vol. 7, pp. 289-296).

[17]     Gomez, R., & Ferguson, M. (2020). Improving self-efficacy for hearing aid self-management: the early delivery of a multimedia-based education programme in first-time hearing aid users. International Journal of Audiology, 59(4), 272-281.

[18]     Gogate, M., Dashtipour, K., Adeel, A., & Hussain, A. (2020). CochleaNet: A robust language-independent audio-visual model for real-time speech enhancement. Information Fusion, 63, 273-285.

[19]     Tammen, M., & Doclo, S. (2022). Deep Multi-Frame MVDR Filtering for Binaural Noise Reduction. arXiv preprint arXiv:2205.08983.

[20]     Nustede, E. J., & Anemüller, J. (2021, August). Towards speech enhancement using a variational U-Net architecture. In 2021 29th European Signal Processing Conference (EUSIPCO) (pp. 481-485). IEEE.

[21]     Grimm, G., Kayser, H., Hendrikse, M., & Hohmann, V. (2018, October). A gaze-based attention model for spatially-aware hearing aids. In Speech Communication; 13th ITG-Symposium (pp. 1-5). VDE.

[22]     Dasenbrock, S., Blum, S., Debener, S., Hohmann, V., & Kayser, H. (2021). A step towards neuro-steered hearing aids: Integrated portable setup for time-synchronized acoustic stimuli presentation and EEG recording. Current Directions in Biomedical Engineering, 7(2), 855-858.

[23]     Dasenbrock, S., Blum, S., Maanen, P., Debener, S., Hohmann, V., & Kayser, H. (2022). Synchronization of ear-EEG and audio streams in a portable research hearing device. Frontiers in Neuroscience, 16.

[24]     Hohmann, V., Paluch, R., Krueger, M., Meis, M., & Grimm, G. (2020). The virtual reality lab: realization and application of virtual sound environments. Ear and Hearing, 41(Suppl 1), 31S.

[25]     Mehra, R., Brimijoin, O., Robinson, P., & Lunner, T. (2020). Potential of augmented reality platforms to improve individual hearing aids and to support more ecologically valid research. Ear and hearing, 41(Suppl 1), 140S.

[26]     Grimm, G., Luberadzka, J., & Hohmann, V. (2019). A Toolbox for Rendering Virtual Acoustic Environments in the Context of Audiology. Acta Acustica United with Acustica, 105(3), 566–578. https://doi.org/10.3813/AAA.919337.

[27]     von Gablenz, P., Kowalk, U., Bitzer, J., Meis, M., & Holube, I. (2021). Individual hearing aid benefit in real life evaluated using ecological momentary assessment. Trends in hearing, 25, 2331216521990288.

[28]     Eastgate, R., Picinali, L., Patel, H., & D'Cruz, M. (2016). 3D games for tuning and learning about hearing aids. The Hearing Journal, 69(4), 30-32.

[29]     Appel, L., Appel, E., Bogler, O., Wiseman, M., Cohen, L., Ein, N., ... & Campos, J. L. (2020). Older adults with cognitive and/or physical impairments can benefit from immersive virtual reality experiences: a feasibility study. Frontiers in medicine, 6, 329.

[30]     Steadman, M. A., Kim, C., Lestang, J. H., Goodman, D. F., & Picinali, L. (2019). Short-term effects of sound localization training in virtual reality. Scientific Reports, 9(1), 1-17.

[31]     Seol, H. Y., & Moon, I. J. (2022). Hearables as a Gateway to Hearing Health Care. Clinical and Experimental Otorhinolaryngology, 15(2), 127.

[32]     Rennies, J. (2023). Better hearing for all: Smart solutions for the clinical, subclinical, and normal-hearing population. In Techniques of Hearing (pp. 77-89). Routledge.

[33]     Blustein, J., Weinstein, B. E., & Chodosh, J. (2022). Over-the-counter hearing aids: What will it mean for older Americans?. Journal of the American Geriatrics Society.

[34]     Edwards, B. (2020, February). Emerging technologies, market segments, and MarkeTrak 10 insights in hearing health technology. In Seminars in Hearing (Vol. 41, No. 01, pp. 037-054). Thieme Medical Publishers.

[35]     Ross, F. (2020). Hearing Aid Accompanying Smartphone Apps in Hearing Healthcare. A Systematic Review. Applied Medical Informatics, 42(4), 189-199.

[36]     Rahme, M., Folkeard, P., & Scollie, S. (2021). Evaluating the accuracy of step tracking and fall detection in the Starkey Livio artificial intelligence hearing aids: A pilot study. American journal of audiology, 30(1), 182-189.